\documentclass[aps,prl,twocolumn,longbibliography,floatfix]{revtex4-2}
\usepackage{amsmath,amssymb,bm,graphicx,hyperref}
\hypersetup{colorlinks=true,linkcolor=blue,citecolor=blue,urlcolor=blue}

\begin{document}
\title{Intrinsic Zak Phase of Solitons: A Geometric Correspondence with the Center-of-Mass Position}
\author{Xiao-Xiao Hu$^{1,3}$}
\email[]{Contact author: huxx2025@lzu.edu.cn}
\author{Chen Cheng$^{1,3}$}
\author{Hong-Gang Luo$^{2,3}$}
\email[]{Contact author: luohg@lzu.edu.cn}

\affiliation{$^{1}$School of Physical Science and Technology, Lanzhou University, Lanzhou 730000, China}
\affiliation{$^{2}$Institute of Fundamental Physics and Quantum Technology, Ningbo University, Ningbo, 315211 China}
\affiliation{$^{3}$Lanzhou Center for Theoretical Physics $\&$ Key Laboratory of Theoretical Physics of Gansu Province, Lanzhou University, Lanzhou 730000, China}
\date{\today}

\begin{abstract}
	The center-of-mass position of a nonlinear soliton in a lattice is commonly attributed to the interplay between nonlinearity and lattice confinement. Here, we reveal its geometric origin by establishing an exact relation between the soliton position and an intrinsic Zak phase. Unlike the Zak phase of a linear Bloch band, this intrinsic Zak phase arises from the nonlinear self-consistent band of the lattice reconstructed by the soliton density. We construct a composite periodic function to define this phase and demonstrate that its topological winding during nonlinear Thouless pumping dictates the quantized soliton displacement. Our results establish band geometry as a fundamental descriptor of nonlinear soliton position in a lattice and provide a geometric framework for their topological transport.
\end{abstract}
\maketitle

\paragraph{Introduction.}
In the modern theory of polarization~\cite{king-smith_theory_1993,vanderbilt_electric_1993,resta_macroscopic_1994,resta_quantum-mechanical_1998,resta_theory_2007}, the Zak phase---the Berry phase accumulated by Bloch states as the crystal momentum sweeps across the Brillouin zone~\cite{zak_berrys_1989}---establishes a fundamental connection between band geometry in momentum space and position in real space~\cite{resta_macroscopic_1994}. It determines the macroscopic polarization of a completely filled fermionic band and the real-space center of a Wannier wave packet~\cite{king-smith_theory_1993,vanderbilt_electric_1993,resta_macroscopic_1994}. When system parameters evolve adiabatically through a cycle, the winding of the Zak phase corresponds to a quantized displacement of the Wannier center and, equivalently, a quantized change in polarization~\cite{resta_macroscopic_1994,thouless_quantization_1983}. This geometric mechanism underlies Thouless pumping, in which band topology governs quantized transport~\cite{thouless_quantization_1983,niu_quantised_1984,xiao_berry_2010,wang_topological_2013,ke_topological_2016,lohse_thouless_2016,nakajima_topological_2016}.

In recent years, research on the Zak phase has expanded beyond single-particle band theory. In interacting many-body systems, interparticle interactions generally destroy the conservation of individual-particle momenta, while the center-of-mass momentum remains conserved in translationally invariant systems~\cite{ke_multiparticle_2017,qin_topological_2018,lin_interaction_2020,lin_topological_2023}. This allows a Zak phase to be defined and related to many-body Wannier centers~\cite{lin_interaction_2020}, providing a geometric description of interaction-induced Thouless pumping~\cite{walter_quantization_2023,huang_interaction-induced_2024,tao_emulating_2025}. The Zak phase has also been studied in nonlinear lattice systems,  including nonlinear waveguide arrays~\cite{szameit_discrete_2024}, nonlinear electrical networks~\cite{hadad_self-induced_2018}, and Bose–Einstein condensates in optical lattices~\cite{watanabe_nonlinear_2016}. In these systems, self-consistent nonlinear Bloch states can be assigned a Zak phase~\cite{bomantara_nonlinear_2017,tuloup_nonlinearity_2020}, extending the geometric description of linear bands to nonlinear states.

Nonlinear lattice systems, however, support not only extended nonlinear Bloch states but also localized solitons~\cite{eiermann_bright_2004,morandotti_observation_2004,zhang_composition_2009,zhang_gap_2009,bersch_optical_2012}. Unlike Wannier states, whose center positions are directly determined by band geometry, nonlinear solitons have profiles and positions shaped by the interplay between nonlinearity and lattice confinement. Previous studies have demonstrated and explored Thouless pumping of nonlinear solitons under adiabatic parameter modulation~\cite{jurgensen_quantized_2021,jurgensen_quantized_2023,di_quantized_2026,jurgensen_chern_2022,fu_nonlinear_2022,fu_two-dimensional_2022,mostaan_quantized_2022,tuloup_breakdown_2023,citro_thouless_2023,hu_pumping_2024,lyu_thouless_2024,cao_nonlinear_2024,cao_transport_2025,kartashov_topological_2025,tao_nonlinearity-induced_2025,tao_anomalous_2026,wang_soliton_2026,bohm_quantum_2026,kartashov_quantized_2026,hu_nonlinearity-induced_2026}. Yet whether the center of a nonlinear soliton in a lattice has a geometric origin remains unknown.

In this Letter, we establish an exact relation between the center-of-mass position of a nonlinear soliton in lattice systems and an intrinsic Zak phase, revealing the geometric origin of the soliton position. Unlike the conventional Zak phase of a linear Bloch band, this intrinsic Zak phase arises from the nonlinear self-consistent band of the lattice reconstructed by the soliton density. By expanding the soliton state in the linear Bloch basis, we construct a composite periodic function whose Berry phase accumulated over the Brillouin zone defines the intrinsic Zak phase without requiring additional gauge fixing. We further show that the winding of this phase during an adiabatic pumping cycle determines the quantized displacement of the soliton. Our results establish self-consistent band geometry as a fundamental descriptor of nonlinear soliton positions and provide a geometric framework for understanding the topological transport of nonlinear localized states.

\paragraph{Model.}
We consider a one-dimensional nonlinear periodic system described by
\begin{equation}\label{eq1}
i\frac{\partial \Psi(x,t)}{\partial t}
=
\left[
-\frac{1}{2}\frac{\partial^2}{\partial x^2}
+V(x)
+g|\Psi(x,t)|^2
\right]\Psi(x,t),
\end{equation}
where $V(x)$ is a periodic lattice potential satisfying $V(x+L)=V(x)$, $L$ is the lattice constant, and $g$ characterizes the strength of the nonlinearity. Here $t$ is the evolution variable, representing time for a Bose–Einstein condensate in an optical lattice, or the propagation distance for light in a periodic nonlinear medium. In the tight-binding regime, Eq.~(\ref{eq1}) reduces to an effective discrete nonlinear lattice model~\cite{alfimov_wannier_2002}, which has been extensively studied in platforms such as optical waveguide arrays and nonlinear circuit networks~\cite{szameit_discrete_2024,hadad_self-induced_2018,di_quantized_2026}. 

Such nonlinear lattice systems support localized stationary solutions (solitons) of the form $\Psi(x,t)=\psi(x)e^{-i\mu t}$, with $\mu$ the chemical potential (or propagation constant) and $\psi(x)$ normalized to unity. Expanding $\psi(x)$ in the Bloch basis $\phi_{nk}(x)=e^{ikx}u_{nk}(x)$ of the linear Hamiltonian $H=-\frac{1}{2}\partial_x^2+V(x)$, where $n$ is the band index, $k$ is the quasimomentum, $u_{nk}(x+L)=u_{nk}(x)$, and $\int_0^L|u_{nk}(x)|^2dx=1$, we obtain
\begin{equation}
\begin{aligned}
	\psi(x)
	&=\frac{L}{2\pi}\int_{\mathrm{BZ}}\sum_nc_n(k)\phi_{nk}(x)\,dk\\
	&=\frac{L}{2\pi}\int_{\mathrm{BZ}}e^{ikx}\mathcal{U}_k(x)\,dk ,
\end{aligned}
\end{equation}
where $c_n(k)=\int_{-\infty}^{\infty}\phi^*_{nk}(x)\psi(x)\,dx$ are the multiband expansion coefficients, the integral runs over the Brillouin zone (BZ), and the composite periodic function is defined as
\begin{equation}
	\mathcal{U}_k(x)=\sum_n c_n(k)u_{nk}(x)
	=e^{-ikx}\sum_m e^{-ikmL}\psi(x+mL),
\end{equation}
where $m\in \mathbb{Z}$ labels the unit cells. Here, $\mathcal{U}_k(x)$ satisfies $\mathcal{U}_{k+G}(x)=e^{-iGx}\mathcal{U}_k(x)$, with $G=2\pi/L$, which connects $\mathcal{U}_k(x)$ across the Brillouin-zone boundary~\cite{my_sm}. Moreover, $\mathcal{U}_k(x)$ shares the unit-cell periodicity of $u_{nk}(x)$, i.e., $\mathcal{U}_k(x)=\mathcal{U}_k(x+L)$, but is generally not normalized: $\langle\mathcal{U}_k|\mathcal{U}_k\rangle=\sum_n|c_n(k)|^2$, with $\langle f|g\rangle=\int_{0}^{L} f^*(x)g(x)\,dx$. Substituting this expansion into Eq.~(\ref{eq1}) yields the equation for $\mathcal{U}_k(x)$~\cite{my_sm}, 
\begin{equation}\label{eq4}
\begin{aligned}
\mu\mathcal{U}_k(x)&=\left[
	\frac{1}{2}(-i\partial_x+k)^2+V(x)
	\right]\mathcal{U}_k(x)\\
&+
	\frac{gL^2}{4\pi^2}\int\int_{\rm BZ}dk_1dk_2
	\mathcal{U}_{k_1}(x)\mathcal{U}_{k_2}^*(x)
	\mathcal{U}_{k+k_2-k_1}(x).
\end{aligned}
\end{equation}
In the composite periodic representation, the nonlinearity acts as a periodic correction that effectively reconstructs the underlying lattice potential $V(x)$. $\mathcal{U}_k(x)$ is therefore the periodic component of the self-consistent solution in the reconstructed lattice. 

Remarkably, the nonlinear eigenvalue $\mu$ is independent of $k$: all quasimomentum components of a given soliton share the same chemical potential (or propagation constant). Consequently, all components have the same dynamical phase, and thus no relative phase difference accumulates during evolution. This prevents the wave packet from spreading due to dispersion, thereby preserving the localized soliton structure. The nonlinear band spectrum of Eq.~(\ref{eq4}) exhibits a flat-band feature, which differs from the bands of extended nonlinear Bloch waves whose eigenvalues vary with $k$~\cite{tuloup_nonlinearity_2020,zhang_composition_2009}. This band flatness originates from the self-consistent structure of the soliton expansion in the Bloch basis, where all quasimomentum components are coupled through the nonlinear term and yield a $k$-independent eigenvalue.

\paragraph{Intrinsic Zak phase.}
Importantly, Eq.~(\ref{eq4}) is invariant under $k\rightarrow k+G$, where $G=2\pi/L$ is the reciprocal lattice vector, while the composite periodic function transforms as $\mathcal{U}_{k+G}(x)=e^{-iGx}\mathcal{U}_k(x)$. This indicates that the Brillouin zone serves as a closed parameter space. Provided that $\langle\mathcal{U}_k|\mathcal{U}_k\rangle>0$ throughout the Brillouin zone, we introduce a Zak phase~\cite{zak_berrys_1989}
\begin{equation}
	\Gamma_{\rm Zak}
	=i\int_{\rm BZ}
	\langle \mathcal{U}_k|\partial_k \mathcal{U}_k\rangle\,dk.
\end{equation}
To preserve the nonlinear self-consistency of Eq.~(\ref{eq4}), $\Gamma_{\rm Zak}$ is constructed from the unnormalized $\mathcal{U}_k$, whose norm $\langle \mathcal{U}_k| \mathcal{U}_k\rangle=\sum_n |c_n(k)|^2$ represents the spectral weight of the soliton in the Bloch basis, reflecting its self-consistent distribution in quasimomentum space. We refer to this spectral-weight-dependent Zak phase as the intrinsic Zak phase of the soliton, which is directly related to its center of mass,
\begin{equation}\label{eq6}
x_c=\int_{-\infty}^{\infty}x|\psi(x)|^2dx=\frac{L}{2\pi}\Gamma_{\rm Zak}.
\end{equation}
This exact relation reveals that the center of mass of a soliton in lattices has a fundamental geometric origin: its center-of-mass position is strictly encoded in the Zak phase of $\mathcal{U}_k$ over the Brillouin zone. This intrinsic Zak phase characterizes the geometry of the nonlinear self-consistent band described by Eq.~(\ref{eq4}).

For soliton branches that preserve spatial inversion symmetry, the intrinsic Zak phase is quantized: $\Gamma_{\rm Zak}$ is restricted to $0$ or $\pi$ modulo $2\pi$~\cite{Asboth2016,Jiao2021,my_sm}. The soliton center of mass can only be located at one of the two inequivalent inversion centers within a unit cell~\cite{my_sm},
\begin{equation}
x_c=0,\quad \frac{L}{2}\quad (\mathrm{mod}\ L).
\end{equation}
Since any other position is not inversion-invariant, these two inversion-invariant positions cannot be connected by continuous deformations that preserve inversion symmetry. This constitutes a topological distinction: solitons located at these two inequivalent inversion centers belong to distinct symmetry-protected classes and cannot be continuously deformed into one another while spatial inversion symmetry is preserved.
\begin{figure*}[htbp]
	\centering
	\includegraphics[width=0.9\linewidth]{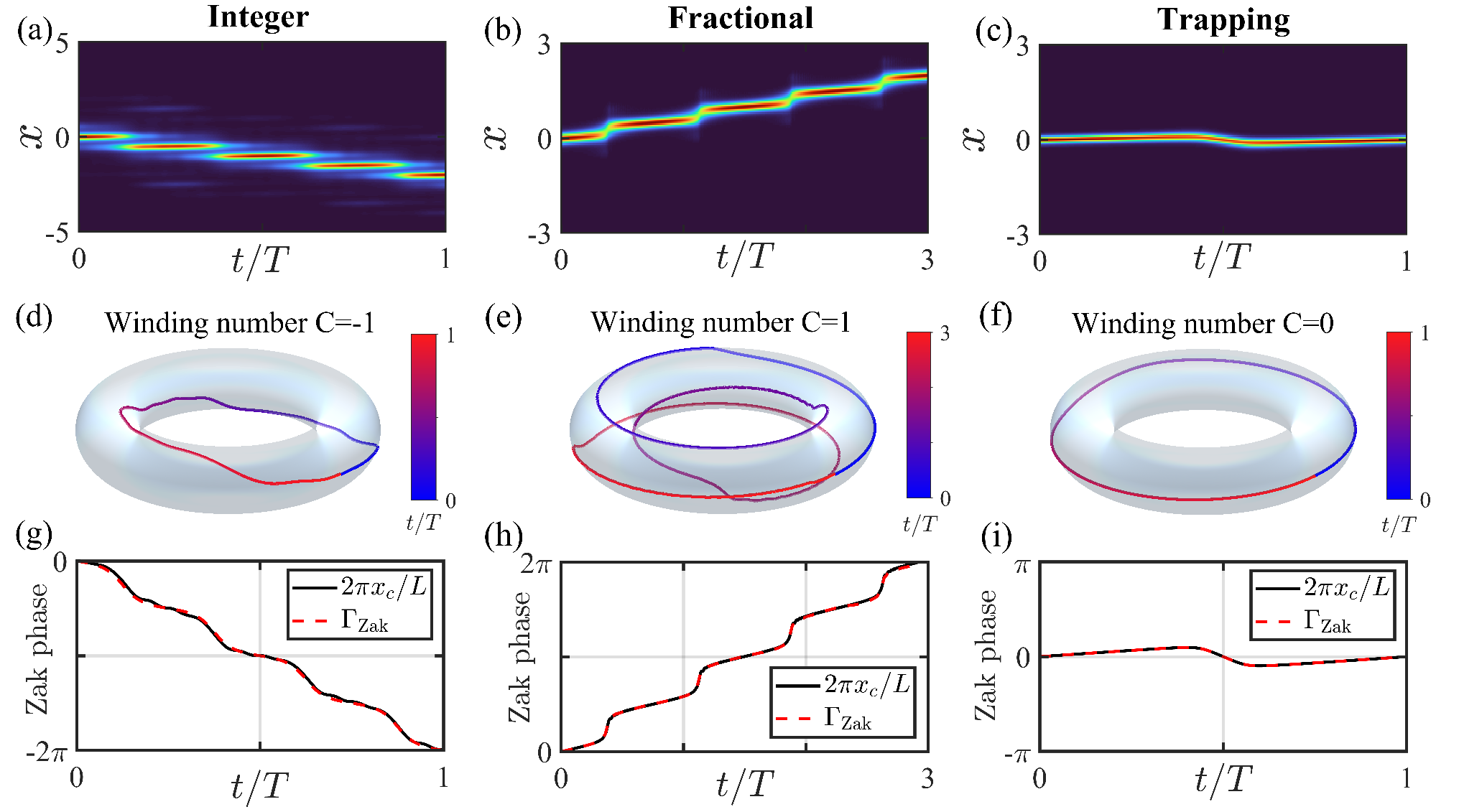}
	\caption{Geometric description of nonlinear Thouless pumping. (a)-(c) Spatiotemporal evolution of the soliton density under three different nonlinear strengths, obtained using the split-step Fourier method: integer pumping ($g=-0.35$), fractional pumping ($g=-7$), and trapping ($g=-20$). (d)-(f) Winding trajectories of the intrinsic Zak phase on the ($k$,$\phi$) parameter-space torus. The corresponding winding numbers are $C=-1$ (left), $C=1$ (middle), and $C=0$ (right). (g)-(i) Comparison between the center-of-mass trajectory $2\pi x_c/L$ (solid black lines) and the intrinsic Zak phase $\Gamma_{\rm Zak}$ (dashed red lines), confirming the center-of-mass–Zak phase correspondence. Other parameters: $p_1=p_2=25$, $v=0.1$, $d_1=1/2$, $d_2=2/3$.}
	\label{fig1}
\end{figure*}
\paragraph{Thouless pumping of solitons.} Having established the exact geometric relation in Eq.~(\ref{eq6}), we now apply it to the adiabatic transport of a nonlinear soliton driven by a slowly varying parameter $\phi(t)$. Provided that the driving rate is much smaller than the characteristic Bogoliubov frequencies of the instantaneous soliton~\cite{liu_nonlinear_2003,pu_adiabatic_2007,liu_berry_2010} and that the nonlinear branch remains dynamically stable and continuously connected~\cite{tuloup_breakdown_2023}, the evolving soliton adiabatically follows the instantaneous nonlinear soliton state~\cite{jurgensen_quantized_2021,jurgensen_quantized_2023,di_quantized_2026}. Consequently, given that its center of mass is related to the intrinsic Zak phase by Eq.~(\ref{eq6}), the center-of-mass displacement along an adiabatic trajectory from $\phi_i$ to $\phi_f$, defined as $\Delta x_c(\phi_f)=x_c(\phi_f)-x_c(\phi_i)$, is given by
\begin{equation}
	\begin{aligned}
	\Delta x_c(\phi_f)
	&=
	\frac{L}{2\pi}
	\left[
	\Gamma_{\rm Zak}(\phi_f)-\Gamma_{\rm Zak}(\phi_i)
	\right]\\
	&=\frac{L}{2\pi}\int_{\phi_i}^{\phi_f}\partial_\phi\Gamma_{\rm Zak}(\phi)d\phi,
	\end{aligned}
\end{equation}
where $\Gamma_{\rm Zak}(\phi)$ is followed continuously along the adiabatic path. For a closed adiabatic cycle of the soliton, its intrinsic Zak phase changes by $2\pi C$, where $C=\frac{1}{2\pi} \oint \partial_\phi\Gamma_{\rm Zak}(\phi)d\phi \in\mathbb{Z}$ is the winding number. This directly yields a quantized soliton center-of-mass displacement,
\begin{equation}
	\Delta x^{\rm cycle}_c
	=
	\frac{L}{2\pi}
	\oint
\partial_\phi\Gamma_{\rm Zak}(\phi)d\phi=CL.
\end{equation}
The soliton undergoes Thouless pumping, with its center-of-mass displacement set by the topological winding of the intrinsic Zak phase. Importantly, the closed cycle of the soliton need not coincide with a single period of the external drive~\cite{jurgensen_quantized_2023,di_quantized_2026,fu_nonlinear_2022}. Let $T$ denote the period of the external modulation and $T_r$ the period over which the soliton returns to a state equivalent to its initial state. Because the soliton is a self-consistent nonlinear state, $T_r$ can differ from $T$. In particular, when $T_r=mT$, the soliton requires $m$ driving cycles to complete one closed cycle. The average displacement per driving cycle is $CL/m$, where $C/m$ can be an integer or a fraction. Thus, when the period of the closed soliton cycle differs from that of the external drive, the soliton can exhibit distinctive fractional Thouless pumping within a single driving cycle~\cite{jurgensen_quantized_2023,fu_nonlinear_2022,hu_nonlinearity-induced_2026}, determined by the winding number $C$ of the intrinsic Zak phase and the number $m$ of  driving cycles per closed soliton cycle.

Without loss of generality, we consider a superlattice potential composed of two commensurate lattice components~\cite{lohse_thouless_2016,nakajima_topological_2016},
\begin{equation}
V(x,\phi)
=
-p_1\cos^2(\frac{\pi x}{d_1})
-p_2\cos^2\!\left(\frac{\pi x}{d_2}+\phi(t)\right),
\end{equation}
where $p_1$ and $p_2$ denote the lattice depths, and $d_1$ and $d_2$ are the spatial periods of the two constituent lattices, satisfying $\frac{d_1}{d_2}=\frac{n_1}{n_2}$, with $n_1$ and $n_2$ being coprime integers. The superlattice period is $L=d_1n_2=d_2n_1$. The relative phase $\phi(t)$ between the two lattices is slowly modulated as $\phi(t)=-vt$, with $v\ll1$, which realizes an adiabatic driving cycle with period $T=\pi/v$. Under this adiabatic drive, the soliton center-of-mass velocity is given by $\dot{x}_c=\frac{L}{2\pi}\partial_\phi\Gamma_{\rm Zak}(\phi)\dot{\phi}$, with its direction of motion determined by the relative sign of $\partial_\phi\Gamma_{\rm Zak}$ and $\dot{\phi}=-v$.
\begin{figure}[htbp]
	\centering
	\includegraphics[width=1\linewidth]{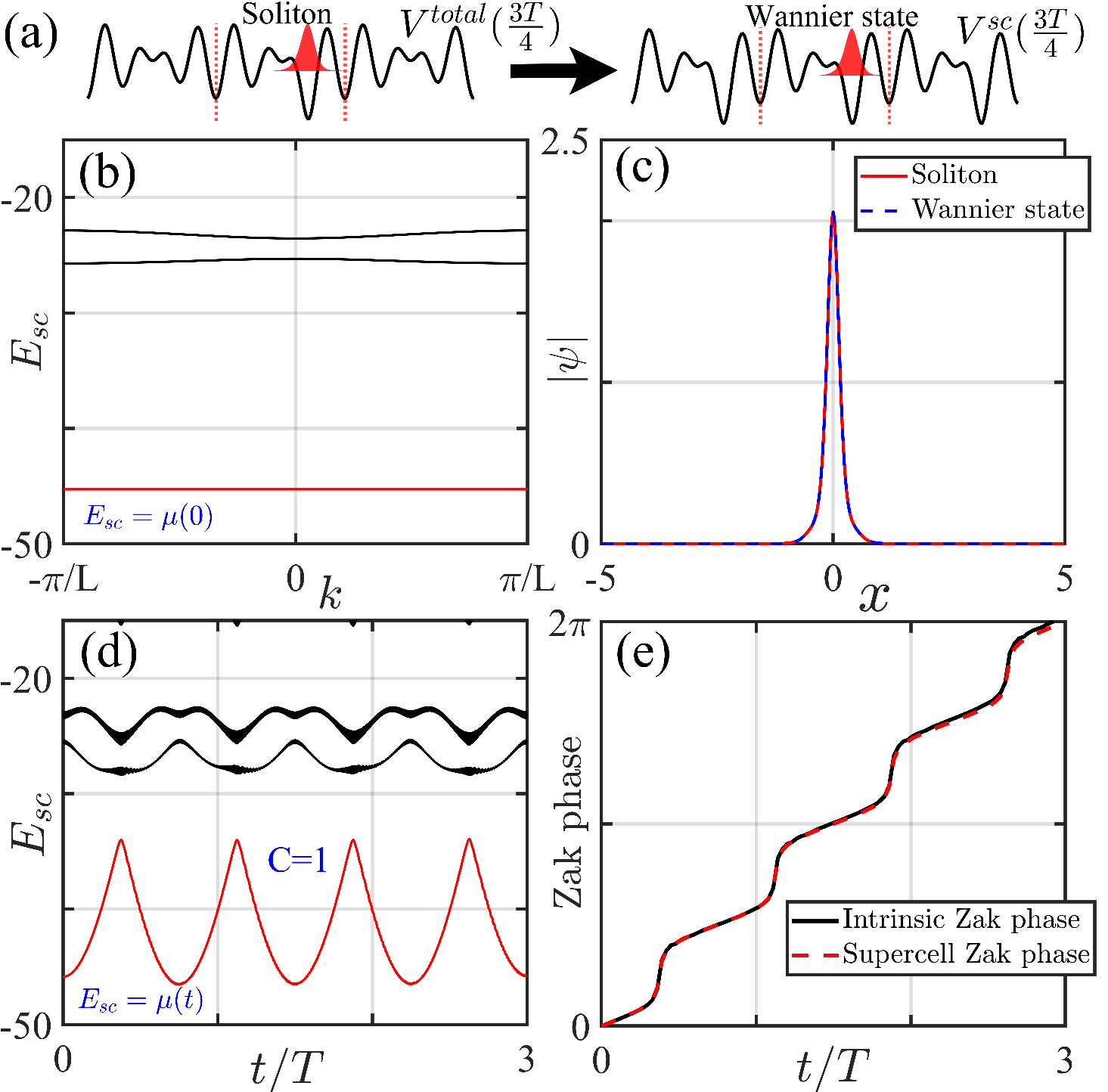}
	\caption{Supercell construction and self-consistent band geometry. (a) Construction of the supercell potential $V^{sc}$. The total effective potential $V^{total}=V+g|\psi|^2$ within the soliton localization region is periodically extended, so that each unit cell reproduces the locally reconstructed potential. (b) Band structure of the supercell Hamiltonian at $t=0$. The lowest band is flat with a $k$-independent energy $E_{sc}=\mu(0)$. (c) Comparison between the soliton density profile (red solid line) and the lowest-band Wannier state (blue dashed line) at $t=0$. (d) Band structure of the time-dependent supercell Hamiltonian. The lowest band exhibits a topological winding number $C=1$, with energy $E_{sc}=\mu(t)$. (e) Comparison between the intrinsic soliton Zak phase (black solid line) and the supercell Zak phase (red dashed line), showing good agreement.}
	\label{fig2}
\end{figure}

During the adiabatic drive, the breaking of spatial inversion symmetry permits continuous motion of the soliton center of mass, with its position determined by the instantaneous value of the intrinsic Zak phase. We consider a gap soliton initially centered at an inversion center of the superlattice~\cite{eiermann_bright_2004,morandotti_observation_2004,zhang_composition_2009,zhang_gap_2009,bersch_optical_2012}. After one complete closed soliton cycle, the superlattice returns to its initial configuration, and the soliton recovers spatial inversion symmetry. Since the intrinsic Zak phase changes by an integer multiple of $2\pi$ over this closed cycle, the soliton at the end of the cycle belongs to the same spatial inversion symmetry-protected class as the initial one. Consequently, the net displacement over a closed soliton cycle can only be an integer multiple of the superlattice period $L$, or the soliton remains trapped at its initial center~[see Fig.~\ref{fig1}(a)-(c)].

Figures.~\ref{fig1}(d)-(i) demonstrate the directional transport of nonlinear solitons under different intrinsic Zak-phase winding configurations. Under a fixed driving protocol, the sign of the winding number $C$ dictates the transport direction, while its magnitude quantizes the net displacement per closed soliton cycle. Figs.~\ref{fig1}(b)(e)(h) illustrate the fractional Thouless pumping, which manifests as a closed soliton cycle spanning multiple driving periods ($T_r=mT$): for $T_r=3T$, the soliton center of mass shifts by approximately $L/3$ per driving period, characterized by an average winding number of $C/m=1/3$.

This distinctive fractional pumping originates from the strong nonlinear reconstruction of the underlying lattice~\cite{tao_nonlinearity-induced_2025,hu_nonlinearity-induced_2026,maisuriya_topologically_2026}, as described by Eq.~(\ref{eq4}). Under strong nonlinearity, the nonlinear coupling term in Eq.~(\ref{eq4}) substantially reconstructs the original potential $V(x,\phi)$. This reconstruction yields a nonlinear self-consistent flat band. Its geometric property is precisely the intrinsic Zak phase, and the topological winding of this phase governs the fractional Thouless pumping of the soliton. Alternatively, this intrinsic band geometry is intuitively revealed via a supercell construction [see Fig.~\ref{fig2}(a)]: the lowest band of the resulting supercell Hamiltonian forms an effective band~\cite{hu_nonlinearity-induced_2026} whose Wannier state reproduces the soliton profile [see Fig.~\ref{fig2}(c)]. This effective band is flat, with a $k$-independent energy $E_{sc}=\mu$, and its Zak phase agrees with the intrinsic Zak phase [see Figs.~\ref{fig2}(b)(d)(e)]. The soliton center of mass is governed by the band geometry described by Eq.~(\ref{eq4}), whose nontrivial topology dictates the topological transport of the soliton. 

Although our analysis is formulated in the Bloch basis of $H=-\frac{1}{2}\partial_x^2+V(x,\phi)$, this choice of basis is not unique. Different choices of the linear Bloch basis may alter the periodicity properties of $\mathcal{U}_k$ and the numerical value of the intrinsic Zak phase, while leaving the form of Eq.~(\ref{eq4}) and the center-of-mass–Zak phase correspondence intact. Furthermore, our approach can be extended to solitons in quasiperiodic lattices, where an appropriate linear basis can be used to characterize both the soliton center of mass and its Thouless pumping~\cite{hu_nonlinearity-induced_2026} in terms of the intrinsic Zak phase. 

\paragraph{Conclusion.}
In summary, this work reveals a fundamental geometric correspondence between the center of mass of a nonlinear soliton and its intrinsic Zak phase in lattices. Under adiabatic driving, the soliton displacement is determined by the change in its intrinsic Zak phase. For a closed pumping cycle, the winding number of this phase determines the quantized displacement of the soliton. Our findings provide a band geometric basis for understanding the position and transport of solitons in lattices and are expected to guide their adiabatic manipulation.
\begin{acknowledgments}
	\paragraph{Acknowledgments.}
	This work is supported by the National Key Research and Development Program of China
	(Grant No. 2022YFA1402704) and the National Natural Science Foundation of China (Grant No. 12247101).
\end{acknowledgments}  
\bibliography{refs} 
\begin{widetext}
  In the Supplemental Material, Section S-1 discusses the gauge freedom of the composite periodic function and shows that the definition of the intrinsic Zak phase of the soliton requires no additional gauge fixing; Section S-2 presents a detailed derivation of the equation for the composite periodic function; and Section S-3 derives the spectral-weight representation of the intrinsic Zak phase and demonstrates its quantization for inversion-symmetric solitons.
\end{widetext}
\setcounter{equation}{0} \setcounter{figure}{0} \setcounter{table}{0} %
\renewcommand{\theequation}{S\arabic{equation}}
\renewcommand{\thefigure}{S\arabic{figure}}
\renewcommand{\bibnumfmt}[1]{[S#1]}
\renewcommand{\citenumfont}[1]{S#1}
\section{S-1 Gauge-invariant definition of the intrinsic Zak phase}
The intrinsic Zak phase of the soliton is defined through the composite periodic function $\mathcal{U}_k(x)=\sum_n c_n(k)u_{nk}(x)$, where $c_n(k)=\int_{-\infty}^{\infty}\phi^*_{nk}(x)\psi(x)\,dx$ are the projection coefficients of the soliton state onto the linear Bloch basis $\phi_{nk}(x)=e^{ikx}u_{nk}(x)$, with the cell-periodic part $u_{nk}(x)$ satisfying the unit-cell normalization $\int_{0}^L|u_{nk}(x)|^2dx=1$. Under an arbitrary band- and momentum-dependent gauge transformation $u_{nk} \to u'_{nk}=e^{i\theta_n(k)}u_{nk}$, since the soliton state $\psi(x)$ is determined, the projection coefficients transform inversely, $c_{n}(k) \to c'_{n}(k)=e^{-i\theta_n(k)}c_{n}(k)$. Therefore, $\mathcal{U}_k(x)$ is independent of the phase-gauge choice of the underlying Bloch basis.

Using the Bloch expansion and partitioning the real-space integral into unit cells, we obtain
\begin{equation}\label{eqs1}
	\begin{aligned}
		\mathcal{U}_{k}(x)&=\sum_n u_{nk}(x)\int_{-\infty}^{\infty}e^{-ikx'}u^*_{nk}(x')\psi(x')\,dx'\\
		         &=\sum_nu_{nk}(x)\sum_m\int_{mL}^{(m+1)L}e^{-iky}u_{nk}^*(y)\psi(y)dy\\
		         &=e^{-ikx}\sum_me^{-ikmL}\psi(x+mL).
	\end{aligned}
\end{equation}
where the last step uses the completeness of the cell-periodic Bloch functions at fixed $k$, $\sum_nu_{nk}(x)u^*_{nk}(y)=\delta(x-y)$. This real-space representation reveals that the $L$-periodicity of $\mathcal{U}_k(x)$ originates from the coherent superposition of the soliton wave function over all lattice translations $mL$ (with $ m \in \mathbb{Z}$). It also determines the transformation of $\mathcal{U}_k(x)$ under a reciprocal-lattice shift $k\to k+G$, with $G=2\pi/L$:
\begin{equation}\label{eqs2}
	\begin{aligned}
		\mathcal{U}_{k+G}(x)
		&=e^{-i(k+G)x}\sum_m e^{-i(k+G)mL}\psi(x+mL)\\
		&=e^{-iGx}\mathcal{U}_k(x),
	\end{aligned}
\end{equation}
where $e^{-iGmL}=e^{-i2\pi m}=1$. This transformation law, $\mathcal{U}_{k+G}(x)=e^{-iGx}\mathcal{U}_k(x)$, naturally connects $\mathcal{U}_k(x)$ across the Brillouin-zone boundary. Consequently, the intrinsic Zak phase can be defined directly from this composite function without requiring any additional gauge fixing.

\section{S-2 Derivation of the equation for the composite periodic function}
We consider a one-dimensional system of $N_x$ unit cells with total length $N_xL$ under periodic boundary conditions. The allowed Bloch momenta are $k_j=2\pi j/(N_xL)$ for $j=0,1,...,N_x-1$. The stationary soliton is expanded as $\psi(x)=\frac{1}{N_x}\sum_{j=0}^{N_x-1}e^{ik_j x}\mathcal{U}_{k_j}(x)$, where $\mathcal{U}_{k_j}(x)=\sum_n c_n(k_j)u_{nk_j}(x)$ is the composite periodic function. The stationary nonlinear Schrödinger equation is 
\begin{equation}\label{eqs3}
	\mu\psi(x)-\left[
	-\frac{1}{2}\frac{\partial^2}{\partial x^2}
	+V(x)
	+g|\psi(x)|^2\right]\psi(x)=0.
\end{equation}
Projecting Eq.~(\ref{eqs3}) onto the momentum component $k_j$, we multiply by $e^{-ik_jx}$ and integrate over the entire ring of length $N_xL$:
\begin{equation}\label{eqs4}
\int_0^{N_xL}dx e^{-ik_jx}\left[\mu\psi(x)-H\psi(x)-g|\psi(x)|^2\psi(x)\right]=0,
\end{equation}
where $H=-\frac{1}{2}\frac{\partial^2}{\partial x^2}+V(x)$ is the linear Hamiltonian. By decomposing the spatial coordinate as $x=x_0+mL$, where $x_0\in [0,L]$ is the intra-cell coordinate and $m$ labels the unit cells, the integration over the full ring splits into a sum over all unit cells and an integral over a single unit cell. Substituting the soliton expansion into Eq.~(S4) and applying the orthogonality relation $\frac{1}{N_x}\sum_{m=0}^{N_x-1}e^{i(k_{j^{\prime}}-k_j)mL}=\delta_{j,j^{\prime}}$, Eq.~(\ref{eqs4}) reduces to
\begin{equation}\label{eqs5}
	\begin{aligned}
0&=\int_0^Ldx_0 \Big[(\mu-H_{k_j})\mathcal{U}_{k_j}(x_0)\\
&-\frac{g}{N_x^2}\sum_{j_1,j_2,j_3=0}^{N_x-1}e^{ipGx_0}\mathcal{U}_{k_{j_1}}(x_0)\mathcal{U}_{k_{j_2}}^*(x_0)\mathcal{U}_{k_{j_3}}(x_0)\Big],
    \end{aligned}
\end{equation}
where $H_{k_j}=\frac{1}{2}(-i\partial_{x_0}+k_j)^2+V(x_0)$, and $G=2\pi/L$ is the reciprocal lattice vector. Here, $k_{j_3}=k_j+k_{j_2}-k_{j_1}+pG$, where $p\in \mathbb{Z}$ is the integer chosen to fold the composite momentum back into the first Brillouin zone. The phase factor is eliminated using the relation $e^{ipGx_0}\mathcal{U}_{k_{j_3}}(x_0)=\mathcal{U}_{k_j+k_{j_2}-k_{j_1}}(x_0)$. Eliminating the sum over $j_3$, Eq.~(\ref{eqs5}) reduces to
\begin{equation}
	\begin{aligned}
	0&=\int_0^Ldx_0\Big[(\mu-H_{k_j})\mathcal{U}_{k_j}(x_0)\\
	&-\frac{g}{N_x^2}\sum_{j_1,j_2=0}^{N_x-1}\mathcal{U}_{k_{j_1}}(x_0)\mathcal{U}_{k_{j_2}}^*(x_0)\mathcal{U}_{k_j+k_{j_2}-k_{j_1}}(x_0)\Big].
    \end{aligned}
\end{equation}
Since Eq.~(\ref{eqs3}) is identically zero at every point, its projection into the momentum component yields an equation that must also be identically zero. We can therefore remove the spatial integration and directly write the equation for $\mathcal{U}_{k}$:
\begin{equation}
	\begin{aligned}
		\mu\mathcal{U}_{k_j}(x)&=\left[
		\frac{1}{2}(-i\partial_x+k_j)^2+V(x)
		\right]\mathcal{U}_{k_j}(x)\\
		&+
    \frac{g}{N^2_x}\sum_{j_1,j_2=0}^{N_x-1}\mathcal{U}_{k_{j_1}}(x)\mathcal{U}^*_{k_{j_2}}(x)\mathcal{U}_{k_j+k_{j_2}-k_{j_1}}(x).
	\end{aligned}
\end{equation}
In the thermodynamic limit ($N_x\to\infty$), the momentum spacing $\Delta k=2\pi/(N_xL)$ approaches zero, and the discrete sums transform into continuous integrals over the first Brillouin zone: $\frac{1}{N_x}\sum_j\to\frac{L}{2\pi}\int_{\mathrm{BZ}}dk$. This rigorously yields the equation for $\mathcal{U}_{k}$ in the main text,
\begin{equation}
	\begin{aligned}
		\mu\mathcal{U}_k(x)&=\left[
		\frac{1}{2}(-i\partial_x+k)^2+V(x)
		\right]\mathcal{U}_k(x)\\
		&+
		\frac{gL^2}{4\pi^2}\int\int_{\rm BZ}dk_1dk_2
		\mathcal{U}_{k_1}(x)\mathcal{U}_{k_2}^*(x)
		\mathcal{U}_{k+k_2-k_1}(x).
	\end{aligned}
\end{equation}
\section{S-3 Spectral weight and inversion-symmetric solitons}
The intrinsic Zak phase in the main text is defined with the unnormalized composite periodic function, $\Gamma_{\rm Zak}=i\int_{\rm BZ} \langle \mathcal{U}_k|\partial_k \mathcal{U}_k\rangle\,dk$. Introducing the spectral weight $\rho(k)=\sum_n|c_n(k)|^2$ and the normalized function $\widetilde{\mathcal{U}}_k=\mathcal{U}_k/\sqrt{\rho(k)}$, we obtain
\begin{equation}
	i\langle \mathcal{U}_k|\partial_k \mathcal{U}_k\rangle=i\rho(k)\langle \widetilde{\mathcal{U}}_k|\partial_k \widetilde{\mathcal{U}}_k\rangle+\frac{i}{2}\partial_k\rho(k).
\end{equation}
Since $\mathcal{U}_{k+G}(x)=e^{-iGx}\mathcal{U}_k(x)$, the spectral weight satisfies $\rho(k+G)=\rho(k)$. Consequently, the contribution involving $\partial_k\rho(k)$ vanishes upon integration over the Brillouin zone, yielding
\begin{equation}\label{eqs10}
	\Gamma_{\rm Zak}=\int_{\rm BZ}\rho(k)\widetilde{A}(k)dk,
\end{equation}
where $\widetilde{A}(k)=\langle \widetilde{\mathcal{U}}_k|i\partial_k \widetilde{\mathcal{U}}_k\rangle$ is the Berry connection. Thus, the intrinsic Zak phase is a spectral-weight-weighted Berry phase. The spectral weight $\rho(k)$ ensures that this geometric phase characterizes the nonlinear self-consistent distribution of the soliton; in the limit of uniform occupation, it naturally recovers the Zak phase of the linear band.

We next consider spatially inversion-symmetric soliton branches. The soliton density satisfies $|\psi(2x_0-x)|^2=|\psi(x)|^2$, where the inversion center of the lattice is restricted to 
\begin{equation}
	x_0=0,\quad \frac{L}{2}\quad (\mathrm{mod}\ L).
\end{equation}
For an inversion-symmetric branch, the soliton wave function satisfies $\psi(2x_0-x)=e^{i\alpha}\psi(x)$. Applying the spatial inversion twice yields  $e^{2i\alpha}=1$, and hence $e^{i\alpha}=\pm1$. With the global phase chosen so that $\psi(x)$ is real, the real-space representation in Eq.~(\ref{eqs1}) gives $\mathcal{U}_{-k}(x)=\mathcal{U}_k^*(x)$, which immediately implies $\rho(-k)=\rho(k)$. Using the inversion symmetry of the soliton wave function, we further obtain $\mathcal{U}_{-k}(x)=e^{2ikx_0}e^{i\alpha}\mathcal{U}_{k}(2x_0-x)$. Consequently, the normalized composite function satisfies $\widetilde{\mathcal{U}}_{-k}(x)=e^{2ikx_0}e^{i\alpha}\widetilde{\mathcal{U}}_{k}(2x_0-x)$ and the Berry connection satisfies $\widetilde{A}(-k)=2x_0-\widetilde{A}(k)$. Combining these inversion constraints with the spectral-weight representation of the intrinsic Zak phase in Eq.~(\ref{eqs10}), we obtain
\begin{equation}
	\Gamma_{\rm Zak}=\frac{1}{2}\int_{BZ}\rho(k)\left[\widetilde{A}(k)+\widetilde{A}(-k)\right]dk=x_0\int_{BZ}\rho(k)dk.
\end{equation}
For a normalized soliton, the total spectral weight over the Brillouin zone is $\int_{\rm BZ}\rho(k)\,dk =G\sum_m\int_0^L|\psi(x+mL)|^2dx=\frac{2\pi}{L}$. It follows that
\begin{equation}
	\Gamma_{\rm Zak}=\frac{2\pi}{L}x_0=0,\quad \pi\quad (\mathrm{mod}\ 2\pi).
\end{equation}
The intrinsic Zak phase takes discrete values, a property constrained by spatial inversion symmetry. Substituting the exact relation $x_c=\frac{L}{2\pi}\Gamma_{\rm Zak}$ from the main text yields
\begin{equation}
	x_c=x_0=0,\quad \frac{L}{2}\quad (\mathrm{mod}\ L).
\end{equation}
Thus, the center of mass of an inversion-symmetric soliton is located at the spatial inversion center of the lattice.
\section{Numerical Methods}
The nonlinear Schr\"{o}dinger equation is solved numerically on a uniform grid with spacing $\Delta x=L_x/N_{grid}$ under periodic boundary conditions $\Psi(x+L_x,t)=\Psi(x,t)$. Typical parameters are the grid size $N_{grid}=2^{12}$, the total system length $L_x=N_xL$, and the time step $\Delta t=0.001$. The initial gap soliton is obtained via Newton's iteration, with the convergence criterion $\max(|\psi_{n+1}-\psi_{n}|)<10^{-12}$. The time evolution is performed using the split-step Fourier method. Because the gap soliton is sufficiently localized and placed far from the boundaries, boundary effects on the dynamics are negligible.

To extract the composite periodic function $\mathcal{U}_k$, we calculate the multiband expansion coefficients $c_n(k_j)$ by projecting the soliton wavefunction onto the linear Bloch basis $\phi_{nk_j}(x)$:
\begin{equation}
	\begin{aligned}
	c_n(k_j)&=\int_{0}^{L_x}\phi^*_{nk_j}(x)\psi(x)\,dx\\
	      &\approx \Delta x\sum_{l=0}^{N_{grid}-1}\phi^*_{nk_j}(x_l)\psi(x_l),
    \end{aligned}
\end{equation}
where $k_j=2\pi j/(L_x)$ $(j=0,...,N_x-1)$ are the crystal momenta, and $x_l=l\Delta x$ denotes the discrete real-space grid points. In practice, the sum over the band index $n$ is truncated to a finite number of bands $M$. The truncation is chosen such that the contribution from higher bands is negligible. 

The composite periodic function can be represented by its coefficient vector $\mathbf{c}(k_j)=[c_1(k_j),c_2(k_j),...,c_M(k_j)]^T$. The intrinsic Zak phase is then evaluated using the discrete Wilson loop:
\begin{equation}
	\begin{aligned}
\Gamma_{\mathrm{Zak}}&=-\arg\left[\prod_{j=0}^{N_x-1}\langle\mathcal{U}_{k_j}|\mathcal{U}_{k_{j+1}}\rangle\right],\\
&=-\arg\left[\prod_{j=0}^{N_x-1}\mathbf{c}^{\dagger}(k_j)\mathbf{c}(k_{j+1})\right]
    \end{aligned}
\end{equation}
where the loop closes via $k_{N_x}\equiv k_0+G$.

All results have been verified to be converged with respect to grid size and time step. For each simulation, two time steps ($\Delta t$ and $\Delta t/2$) and two spatial steps ($\Delta x$ and $\Delta x/2$) are used. The results at the same $t$ are compared; if the peak amplitude difference of the wave function is below $10^{-8}$, the results are accepted. Otherwise, the step sizes are further halved and the simulation repeated. 
\end{document}